\documentclass[preprint,12pt]{elsarticle}

\usepackage{amssymb}
\usepackage{amsthm}
\usepackage{url}
\usepackage{verbatim}
\graphicspath{ {images/} }
\usepackage[dvips]{color}

\usepackage{lineno}

\journal{Experimental Thermal and Fluid Science}

\begin{document}
\begin{frontmatter}



\title{Multiple vortex structures in the wake of a rectangular winglet in ground effect}


\author{Clara M. Velte}
\address{Department of Mechanical Engineering, Technical University of Denmark, 2800 Kgs. Lyngby, Denmark}

\author{Martin O.L. Hansen}
\address{Department of Wind Energy, Technical University of Denmark, 2800 Kgs. Lyngby, Denmark}
\address{Centre for Ships and Ocean Structures, Norwegian University of Science and Technology, 7491 Trondheim, Norway}

\author{Valery L. Okulov}
\address{Department of Wind Energy, Technical University of Denmark, 2800 Kgs. Lyngby, Denmark}
\address{Institute of Thermophysics, Siberian Branch of the Russian Academy of Sciences, 630090 Novosibirsk,
Russia
}

\begin{abstract}
Patterns of vorticity in the wake of a single rectangular winglet (vortex generator) embedded in a turbulent boundary layer have been studied using Stereoscopic Particle Image Velocimetry (SPIV). The winglet was mounted normally to a flat surface with an angle to the oncoming flow. A parametric study varying the winglet height (constant aspect ratio) and angle has shown, contrary to the common classical single tip-vortex conception, that the wake generally consists of a complex system of multiple vortex structures. The primary vortex has previously been discovered to contain a direct coupling between the axial and the rotational flow. In the current work, even the longitudinal secondary structures detected  from measured streamwise vorticity display similar behavior. A regime map depicting the observed stable far wake states of the multiple vortices as a function of winglet height and angle reveals complex patterns of the flow topologies not only with the primary tip vortex, but with the additional secondary structures as well. A bifurcation diagram shows distinct regimes of the various secondary structures as well as how the primary vortex is in some cases significantly affected by their presence. These data should serve as inspiration in the process of generating longitudinal vortices for enhancement of heat and mass transfer in industrial devices since the multiple vortex regimes can help improve the conditions for these exchanges. Further, these results point to a weakness in existing inviscid models not accounting for the possibility of multiple vortical structures in the wake.
\end{abstract}


\end{frontmatter}


\section{Introduction}
\label{sec:Introduction}

Vortex generators were described by Taylor as early as in the late 1940s~\cite{Taylor1947,Taylor1948a,Taylor1948b,Taylor1950} and have for many years been used to, e.g., delay separation and enhance mixing of momentum and heat. A common strategy to increase the heat transfer coefficient is to generate longitudinal vortices in flows whose rotating motion induces an increasing exchange of hot and cold fluid, see, e.g.,~\cite{JacobiShah1995,GentryJacobi1997,Fiebig1991,Tiggelbeck1992,Hochdorfer1995,Min2012}. Min and co-workers~\cite{Min2010} have observed that the topology and composition of the longitudinal vortices can be of uttermost importance to heat transfer enhancement. Previous research has put a great deal of effort into trying to understand and optimize the effect of the devices without succeeding in finding a general approach valid independently of test bed. Except for the comprehensive combined theoretical and experimental summary of Pearcey~\cite{Pearcey1961}, most early studies use surface visualizations and measurement techniques providing integral measures (e.g., integral forces and observation of reversed flow) in specific applications rather than seeking understanding in the detailed flow physics and characteristics. The work of Lin and co-workers presents many generic experimental studies on VGs for separation control~\cite{Lin1989,Lin1990,Lin1991,Lin1999,Lin2002}. One of their many findings is that for turbulent boundary layers, the smaller micro-vortex generators work very well in turbulent boundary layers even if their height is only a fraction of the boundary layer height. Their high efficiency is, from actual measurement results, explained by the much fuller velocity profile of a turbulent boundary layer as compared to a laminar one and are also hypothesized to function in themselves like `turbulators' rather than like the classical picture of a single wing tip vortex provided by Taylor~\cite{Taylor1947,Taylor1948a,Taylor1948b,Taylor1950}.

Proper investigations of the complete wake behind vortex generators have only recently become practically possible with the advent of optical techniques such as digital Particle Image Velocimetry (PIV). Most of   these studies investigate the velocity flow field behind rows of vortex generators producing co- or counter-rotating vortex cascades for certain applications. In a recent study~\cite{Velteetal2009} investigating   snapshots of the three-component velocity field and stream-wise vorticity field behind a single rectangular vortex generator, a secondary structure was discovered which was not previously observed in studied cascade configurations (see, e.g.,~\cite{Lau1995,GodardStanislas2006}). This secondary vortex has shown to be able to substantially perturb the primary vortex core itself~\cite{Velteetal2009} as well as its position~\cite{Orlandi1990,Velte2012}. Its generation was not fully understood in the setting with an oncoming stream-wise boundary layer and therefore naturally triggered the present study. Previous studies have revealed similar structures formed by local separation of the boundary layer in the lateral direction by the pressure gradient imposed by the primary vortex on the wall. Known examples include simulated aircraft trailing vortices by Harvey \& Perry~\cite{HarveyPerry1971} as well as in two-dimensional computations of vortex pairs approaching a wall (see, e.g.,~\cite{Orlandi1990}), 2D visualizations~\cite{Tiggelbeck1992} and in three-dimensional flow simulations (see, e.g.,~\cite{LutonRagab1997}). Despite this prior knowledge, it is interesting to observe that none of the previous studies specific to vortex generators, at least known to the authors, report about this structure until only recently~\cite{Velteetal2009}.

In addition, the pressure distribution closest to the wall around the leading edge of the vane in the lowest part of the boundary layer results in a horseshoe vortex system, see~\cite{Velte2012}. This is very similar to other types of junction flows around, e.g., cylinders~\cite{SumerFredsoe2006}, blades of axial turbines~\cite{Langston2001} and wing-body junctions~\cite{Simpson2001}. These structures have also been observed and described around wedge-type vortex generators~\cite{Terzis2015}. This effect was difficult to capture in the PIV measurements in the current work, since one sleeve of the horseshoe vortex was swept under the primary vortex and joined the local separation (which has the same sign of rotation) at a very early stage of the wake development. Due to problems of reflections from surfaces, such as the vortex generator trailing edge, one cannot always measure close enough to the generator to capture this vortex sleeve. Water channel visualizations could therefore aid in visualizing and confirming this anticipated effect, see~\cite{Velte2012}.

The separate existence of these two types of flow mechanisms in the particular setting of vortex generator flows is now established~\cite{Velteetal2009,Velte2012}, but the resulting far wake has not previously been studied. The present work therefore aims to extend the experimental investigation by studying the combination of these non-linear viscid flow mechanisms, showing that the resulting flow field can display a high degree of complexity as opposed to what has previously been assumed. Further, the existence of the observed viscid secondary structures and their influence on the main vortex are not previously accounted for in existing (usually inviscid) analytical models of the VG wake. This simplification can potentially yield considerable deviations of these models from the actual flow, which stresses the motivation for the current study. The focus of this study is the practically important far wake, which is displaying a steady wake topology from about 5 vortex generator heights downstream of the vane and on. The unsteady developing near wake appears only to be of minor practical interest since the performance of the vortex generator is mainly concentrated to the far wake (see, e.g.,~\cite{GodardStanislas2006}). In the present work, the regimes of the far wake are therefore mapped as a function of vortex generator angle and height as well as the dependency of the circulation of the primary vortex to its proximity to the wall.

\section{Experimental Method}
\label{sec:ExperimentalMethod}

\begin{figure}[!ht]
\includegraphics[width=0.8\textwidth]{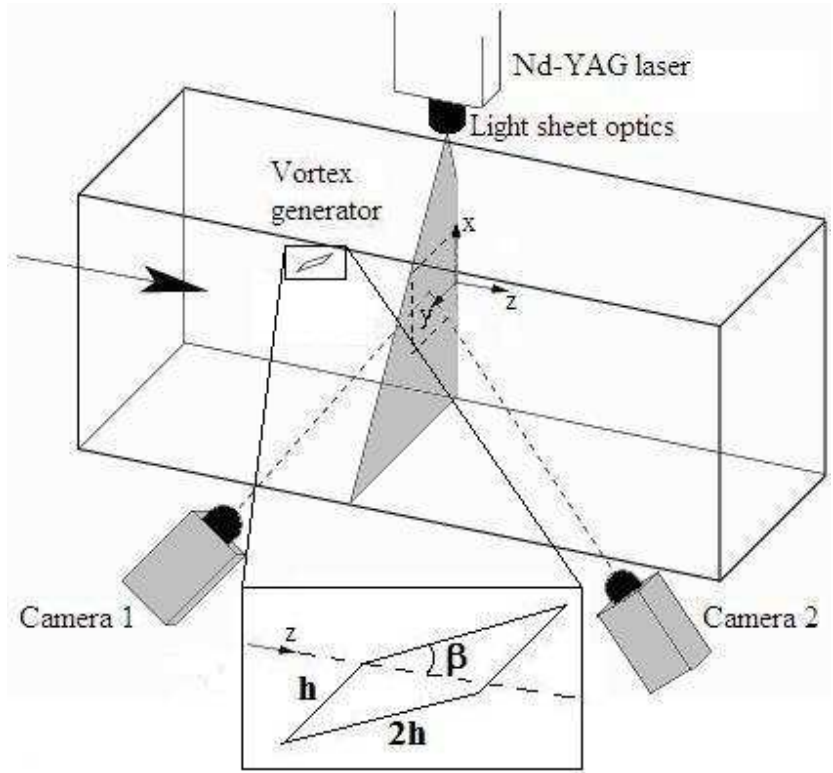}
\caption{Sketch of the wind tunnel measurement setup (from~\cite{Velteetal2009}).}
\label{fig:1}
\end{figure}

\begin{figure}[!ht]
\includegraphics[width=1.0\textwidth]{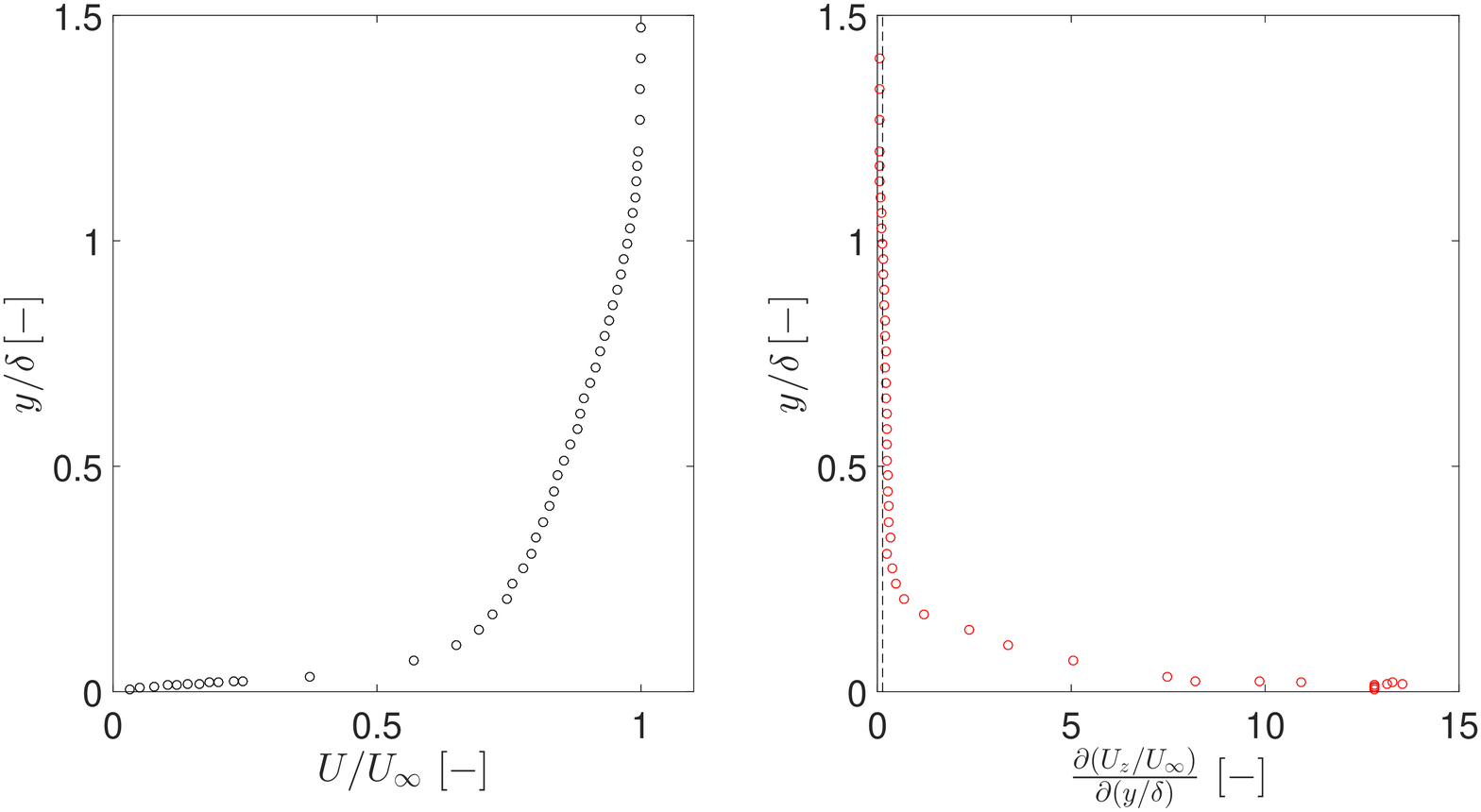}
\caption{Streamwise time averaged velocity profile measured using LDA (left) and its corresponding spatial derivative in the wall-normal direction as an estimate of the spanwise vorticity (right). The dashed line in the right graph indicates $1\%$ of the maximum spanwise vorticity. The velocity is normalized by the free stream velocity and the y-axis by the local boundary layer thickness $\delta$.}
\label{fig:2}
\end{figure}

Stereoscopic Particle Image Velocimetry (SPIV) experiments were carried out in a closed-circuit wind tunnel as described in~\cite{Velte2012} and in more detail in~\cite{Velteetal2009}. The measurement setup is sketched in Figure~\ref{fig:1}. The closed-loop wind tunnel, with cross section $300 \times 600\, mm$, test section length $2\, m$ and a contraction ratio of $8$:$1$, was set to run at a free stream velocity of $U_{\infty} = 1.0\, ms^{-1}$, corresponding to a Reynolds number $Re_{\delta} = U_{\infty}\delta/\nu = 1670$ based on the boundary layer thickness. The height of the largest vortex generator was set to the boundary layer thickness $\delta = 25\, mm$ at the position of the vortex generator. The wind tunnel speed was measured from the pressure drop across an orifice plate. At the inlet a turbulence generating grid with mesh size $39\, mm$ was positioned, producing a turbulent free-stream and boundary layer. The turbulence intensity at the inlet has from LDA measurements been found to be $13\%$~\cite{Velte2009}. The boundary layer thickness was measured using both SPIV and laser Doppler anemometry (LDA). The time averaged streamwise velocity profile normalized by the free stream velocity measured by LDA is displayed in Figure~\ref{fig:2}. Due to the concave shape of the velocity profile observed from the PIV measurements covering a larger wall-normal distance (not shown), the boundary layer thickness was determined by estimating the vorticity from the LDA profile, since the boundary layer can be considered the (viscid) vorticity contained part of the flow. An analogue definition could therefore be to define the boundary layer   thickness as the distance over which the spanwise vorticity has been reduced to $1\%$ of the maximum value. Since, by an order of magnitude analysis, one can discard the first term in the first vorticity component, one can estimate the spanwise vorticity satisfactorily by only the streamwise velocity and the wall-normal coordinate as also shown in Figure~\ref{fig:2} (see, e.g.,~\cite{Velte2008}).

The vortex generators are rectangular plates of heights $h=5$, $10$, $15$, $20$ and $25\, mm$ and a thickness of about $0.5\, mm$. The vane length is always set to $l=2h$ so that the aspect ratio is kept constant. The vortex generator was always positioned $750\, mm$ downstream of the inlet in the center of the test section on the widest wall. A sketch of the wind tunnel test section taken from~\cite{Velteetal2009} is shown in Figure~\ref{fig:1}. The coordinate system is defined in the Figure~\ref{fig:1}. $z$ is the axial flow direction, $y$ is the wall-normal direction and $x$ is the spanwise direction. To accurately set the vortex generator angle, $\beta$, the vortex generator was attached to a pin which could be accessed from outside the test section through a hole in the test section wall. This pin was in turn attached to a pointer arm placed over a protractor indicating the relative angle of the actuator to the mean flow direction. The protractor had a radius of $200\, mm$ and grading for integer values of each degree. The device angle of incidence $\beta$ could therefore be set with a high accuracy.

The measurements were carried out in spanwise planes perpendicular to the main flow direction downstream of a single rectangular vane (see Figure~\ref{fig:1}) to obtain reliable values of the streamwise velocity and the streamwise vorticity throughout the wake. The stable far wake ($\sim 5h$ and on) was mapped in a parametric study, varying the vortex generator angle to the incoming flow ($\beta = 9 - 54^{\circ}$ with $3^{\circ}$ spacing) and the device height. In this manner, the various regimes of the wake can be mapped as a function of the vane parameters (angle and height). For a better understanding of the emergence of these states, the downstream evolution of each of these has been investigated by measurements in a number of positions downstream of the vane ($z/h = 0.25, \,1,\, 2,\, 3,\, 5,\, 8$ and $10$), covering both near and far wake. For convenience the SPIV system was rigidly mounted on a traversing system, enabling measurements in varying streamwise positions without requiring new calibrations at each new position.

The SPIV equipment included a double cavity NewWave Solo 120 XT Nd-YAG laser (wavelength $532\, nm$) capable of delivering light pulses of $120\, mJ$. The pulse width, i.e., the duration of each illumination pulse, was $10\, ns$. The light-sheet thickness at the measurement position was $2 \,mm$ and was created using a combination of a spherical convex and a cylindrical concave lens. The equipment also included two Dantec Dynamics HiSense MkII cameras ($1344 \times 1024$ pixels) equipped with $60\, mm$ lenses and filters designed to only pass light with wavelengths close to that of the laser light. Both cameras were mounted on Scheimpflug angle adjustable mountings. The seeding, consisting of DEHS (diethyl-hexyl-sebacin-esther) droplets with a diameter of $2-3\,\mu m$, was added to the flow downstream of the test section in the closed-circuit wind tunnel in order to facilitate a homogeneous distribution of the particles before they enter the test section. The laser was placed above the test section, illuminating a plane normal to the test section walls (see Figure~\ref{fig:1}). The two cameras were placed in the forward scattering direction. The angle of each respective camera to the laser sheet was $45^{\circ}$. The f-numbers of the cameras were set to $2.8$, yielding a depth of field which is small but sufficient to cover the thickness of the laser sheet and keeping all illuminated particles in focus while still attaining sufficient scattered light from the tracer particles.

In order to avoid reflections from the wall and the vortex generator within the wavelength band of the camera filters, these areas were treated with a fluorescent dye, Rhodamine $6G$, mixed with matt varnish to obtain a smooth surface and to ensure that the dye stayed attached. A calibration target was aligned with the laser sheet. This target had a well-defined pattern, which could be registered by the two cameras to obtain the geometrical information required for reconstructing the velocity vectors received from each camera to obtain a full description of all three velocity components in the plane. Calibration images were recorded with both cameras at five well defined streamwise positions throughout the depth of the laser sheet in order to capture the out-of-plane component in the reconstructed coordinate system of the measurement plane under consideration. A linear transform was applied to these images for each camera to perform the reconstruction. This procedure was executed both previous to and after the conduction of the measurements to ensure that no drift had occurred.

The images were processed using Dantec Dynamic Studio software version $2.0$. Adaptive correlation was applied using refinement with an interrogation area size of $32 \times 32$ pixels. Local median validation was used in the immediate vicinity of each interrogation area to remove spurious vectors between each refinement step. The overlap between interrogation areas was $50\%$. For each measurement position, $2\,000$ realizations were acquired. As a check of statistical convergence, the average velocity maps were computed for $500$, $1\,000$, $1\,500$ and $2\,000$ samples, displaying the same topology. The recording of image maps was done with  an acquisition rate of $2.0\, Hz$, ensuring statistically independent realizations based on the convection velocity $U_{\infty} = 1.0\, ms^{-1}$ and the mesh size $d = 0.039 \,m$, yielding a time scale of $t = d/U_{\infty} = 0.039\, s$. The velocity vector maps contain $73 \times 61$ vectors.

The linear dimensions of the interrogation areas ($\Delta x, \Delta y$) = ($1.55,1.04$) $mm$ can be compared to the Taylor microscale and the Kolmogorov length scale estimated to $\lambda_f \approx 9\, mm$ and $\eta \approx 0.5\, mm$ from LDA measurements~\cite{Velte2009}, which ensures sufficient spatial resolution in relation to the scales of the measured flow. Furthermore, it was possible to align the calibration target to the laser sheet with an accuracy significantly smaller than the measured Taylor microscale at that position. The disparity between the calibration target and the light sheet was typically around $0.05$ pixels, i.e., smaller than the optimal measurement accuracy of the PIV system ($0.1$ pixels). This corresponds to a particle displacement error of about $2\,\mu m$, which can be considered negligible in comparison to the scales in the measured flow.

\section{Development of vortex system behind vortex generator}
\label{sec:Developmentofvortexsystembehindvortexgenerator}

\begin{figure}[!ht]
\includegraphics[width=0.7\textwidth]{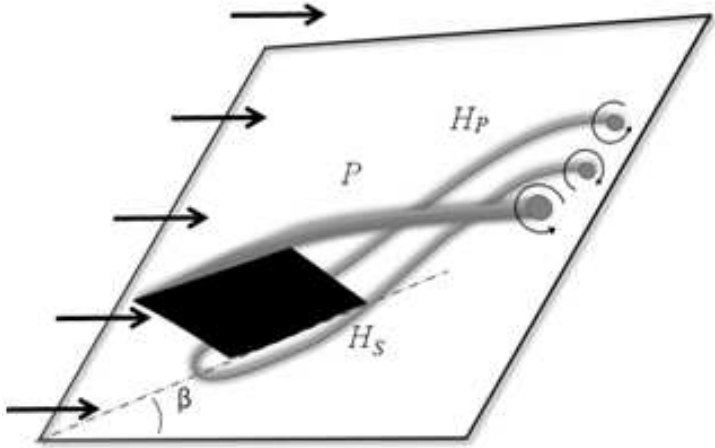}
\caption{Topology of the basic vortex structure found from dye visualizations (figure from~\cite{Velte2012}).}
\label{fig:3}
\end{figure}

\begin{figure}[!ht]
\begin{minipage}{.48\textwidth}
\includegraphics[width=\textwidth]{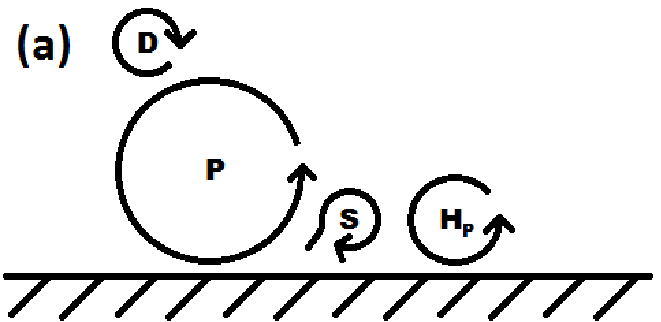}
\end{minipage}
\begin{minipage}{.48\textwidth}
\includegraphics[width=\textwidth]{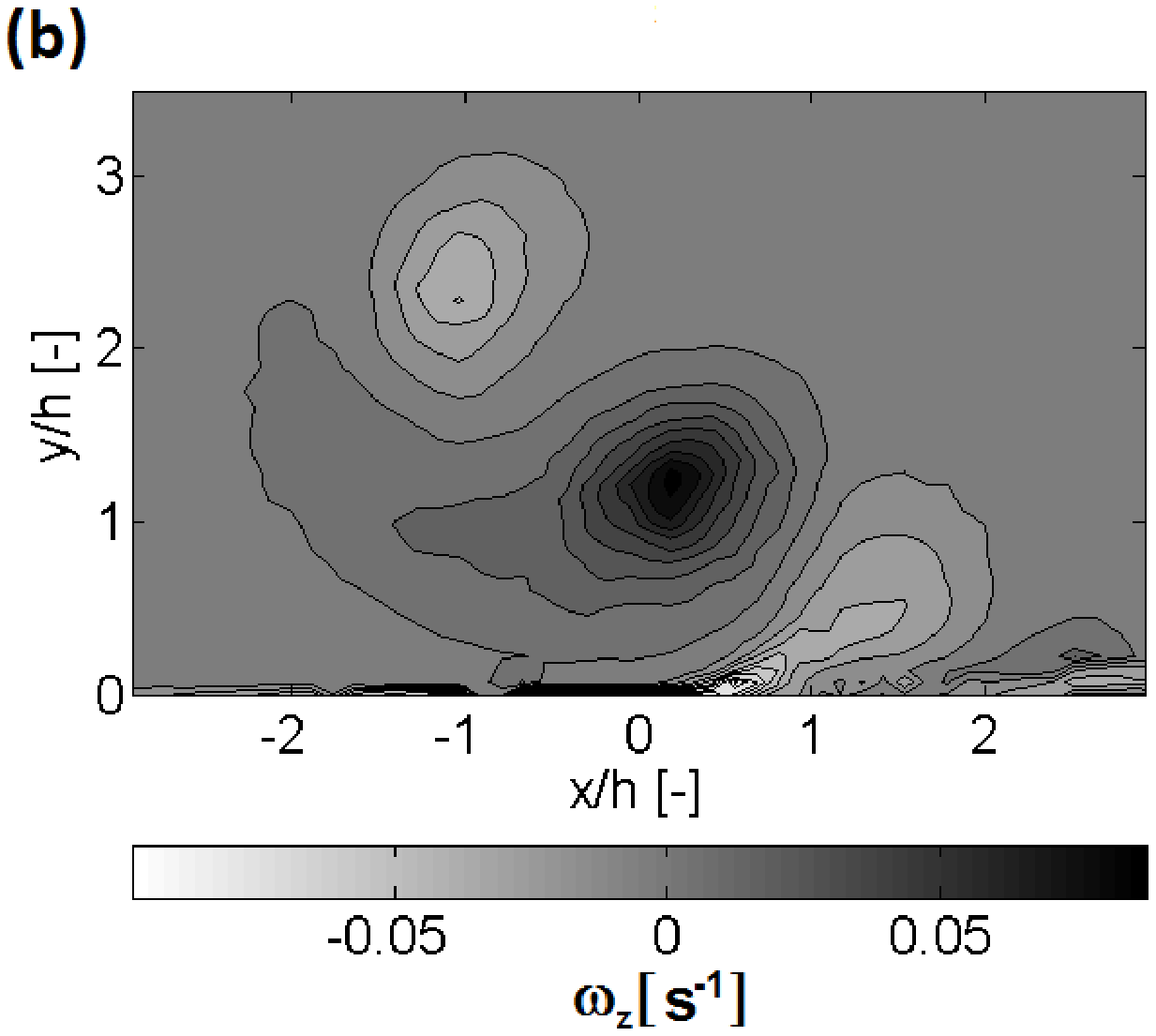}
\end{minipage}
\caption{Vortex structure topology behind the vortex generator (a) by direct combination of the basic vortex system and the secondary vortex structure and (b) the corresponding observed far wake state ($z/h=10$)  for $\beta = 33^{\circ}$ and $h/\delta = 1$.}
\label{fig:4}
\end{figure}

The main study is performed in a wind tunnel using Stereoscopic PIV. These measurements show a complex vortex system which is composed of two basic mechanisms:

\begin{itemize}
\item A \textit{\textbf{basic vortex system}} constituted by the primary (wing tip) vortex and a horseshoe vortex generated from the roll‐up vortex around the leading edge of the vane, see Figure~\ref{fig:3}.
\item A \textit{\textbf{secondary vortex structure}} created by local separation of the boundary layer in the lateral direction between the primary vortex and the wall, as can be seen, e.g., in Figure~\ref{fig:4}.
\end{itemize}

These mechanisms will in the following first be described independently. We will then continue by describing the observations of the measured wake structures, showing the \textit{combined} effect of these two flow mechanisms.

\subsection{Description of the basic vortex system}
\label{sec:Descriptionofthebasicvortexsystem}

In previous water channel experiments~\cite{Velte2012}, prior knowledge about these basic vortices and how they are generated facilitated the use of dye visualizations through pigment injection, since this revealed the generation area of each vortex and therefore also the ideal dye injection points. From the dye visualizations, the flow topology of the basic vortex system was visualized and is reproduced in the sketch in Figure~\ref{fig:3}. Each vortex is indicated in the figure, $P$ being the primary one and $H_P$ and $H_S$ the pressure side and suction side horseshoe vortices, respectively. For the vortex generator, due to the generation of the strong primary tip vortex, the flow will appear somewhat differently to that generated by axial gas turbine blades~\cite{Langston2001}. The primary vortex, $P$, will, at a very early stage of its generation, sweep the suction side horseshoe vortex of opposite sign, $H_S$, underneath it. The pressure side horseshoe vortex, $H_P$, usually seems to evolve undisturbed by the rest of the vortex system throughout the observed downstream range. It should be noted that sometimes this disturbance of the boundary layer introduced by the vane is not strong enough, resulting in that both sleeves of the horseshoe vortex cannot be identified with separate vortex cores and then instead appear as vorticity track sticking to the tunnel or channel wall. Thus, two basic structures consisting of one ($P$) or three ($P$, $H_S$ and $H_P$) vortices will typically emerge in the wake behind the vane.

\subsection{The separated secondary vortex structure(s)}
\label{sec:Theseparatedsecondaryvortex structure(s)}


Figure~\ref{fig:4}$a$ displays the expected basic topology of the two combined vortex systems while Figure~\ref{fig:4}$b$ displays a measured streamwise vorticity plot with corresponding vortex structures. Due to the sweeping of $H_S$ in under $P$, the $H_S$ structure is expected to merge with $S$, adding to its strength. Though this structure is really a combination of the two, the notation $S$ will be used for this structure in the remainder of the article (see Figure~\ref{fig:4}$a$). As the primary vortex (up to a certain point) becomes increasingly stronger, the continuously generated separation region grows to eventually detach and form a discrete vortex, here denoted $D$.

\subsection{Measured vortex structure regimes in the wake}
\label{sec:Measuredvortexstructureregimesinthewake}

The subsequent motion of the vortices of the combined basic and secondary systems is clearly a complex 3D vortex dynamics problem. Our parametric study shows that direct superposition of these two mechanisms (such as that in Figure~\ref{fig:4}) is merely one of many possible states. By varying geometrical parameters such as vane angle and height, we have observed as many as seven different states.

\begin{figure}[!ht]
\includegraphics[width=0.88\textwidth]{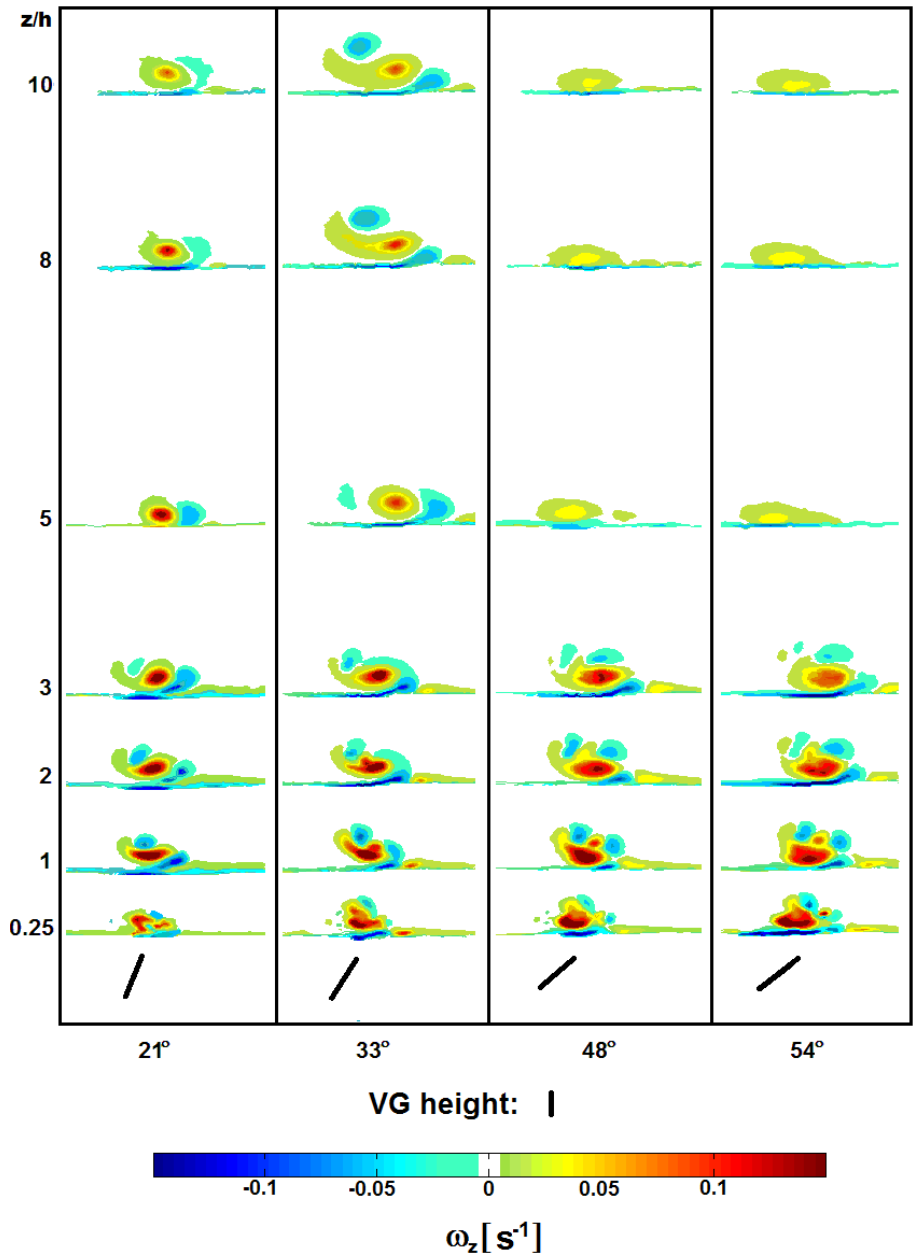}
\caption{Downstream development of the vortex generator wake for $h/\delta = 1$. The vortices are displayed by iso-contours of the streamwise vorticity with a threshold level (white) to filter out the effects of measurement noise.}
\label{fig:6}
\end{figure}

Four of the most dominantly occurring measured states and their rough overall topology development throughout the wake in the form of streamwise vorticity plots are illustrated in Figure~\ref{fig:6}. To ensure that measurement noise does not affect the results, a threshold criterion of the vorticity with a minimum value equal to twice the measurement noise level has been used to detect the existence of vortices. Values below this threshold are filtered out and displayed in white. Note that this rough sketch has been added only to demonstrate the development stages of the vortex system behind the vortex generator. The details of the specific cross-sections of the wake will be considered below.

From visual inspection, the wake can be divided into near wake ($\sim 0-5h$) and far wake (after $\sim 5h$ and on), where the near wake is clearly a complex developing region which stabilizes into a steady topology in the far wake. In practical applications, vortex generators are commonly placed a certain distance upstream of where they are expected to impact the flow (see, e.g.,~\cite{Lin2002,GodardStanislas2006,Velte2009,Velte2008}). This distance corresponds well to the observed near wake region. Since only the far wake is of practical interest, the complex unsteady near wake is not investigated here in detail.

The state depicted in Figure~\ref{fig:4}$a$ is represented in the $\beta = 33^{\circ}$ case in Figure~\ref{fig:6}. The existence of $H_P$ is also observed for $\beta = 48^{\circ}$, but without $H_S$ or $D$ present. The absence of the second sleeve of the horseshoe vortex, $H_S$, merging with $S$, may be explained in that the primary vortex $P$ was too close to the wall to be able to sweep $H_S$ in underneath itself. Instead, the main vortex spreads the opposite sign vorticity of the $H_S$ vortex along the wall and therefore not easily detectable by the SPIV measurement technique. One  relatively common state, where all three basic vortices ($P$, $H_P$ and $H_S$) behind the vane are well pronounced, is reproduced for $\beta = 21^{\circ}$, but $H_P$ is weak in the far wake. This case coincides with the one in the topological sketch in Figure~\ref{fig:3}. Finally, for a very large vane angle, $\beta = 54^{\circ}$, the measured average streamwise vorticity in Figure~\ref{fig:6} displays only the primary vortex $P$, which has long been assumed to be the standard wake   produced by a vortex generator.

\begin{figure}[!ht]
\includegraphics[width=0.9\textwidth]{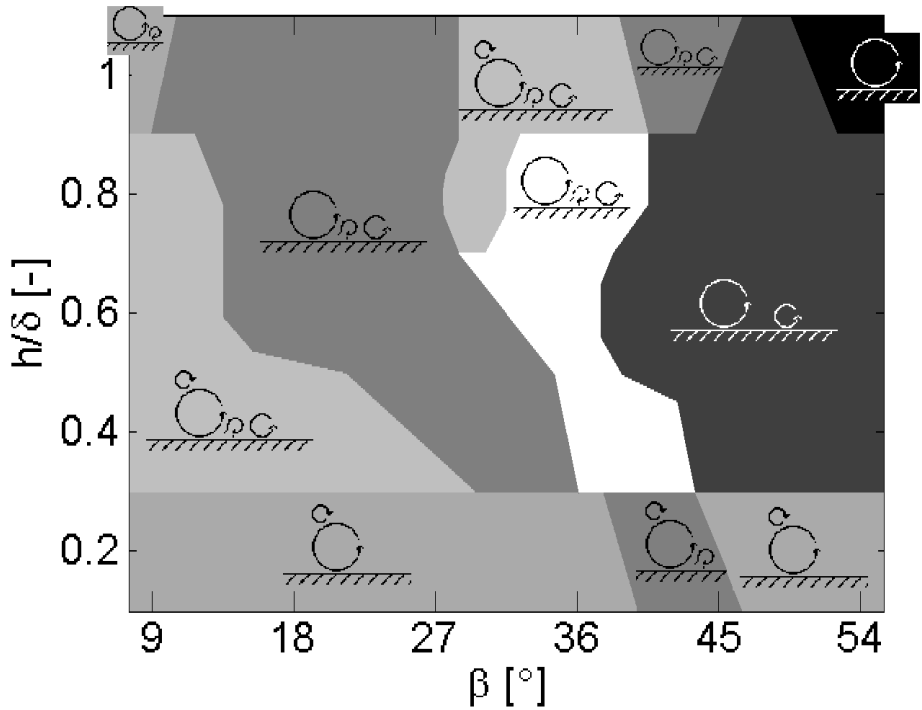}
\caption{Regime map of the different vortex structures produced in the far wake ($z/h = 10$) behind the vortex generator mapped across vane height/boundary layer thickness and vane angle space.}
\label{fig:7}
\end{figure}

From the parametric study, the stable states in the far wake are mapped for different vane angles, $\beta$, and vane heights relatively to the boundary layer thickness, $h/\delta$. The resulting regimes produced by the actuator are displayed in Figure~\ref{fig:7}. As is obvious from the figure, the resulting far wake is very much dependent on the physical parameters of the vane and the oncoming flow.

For the smallest vanes, one of the most evident features is that no horseshoe vortex is present. One plausible cause for its absence is that the vane height is small compared to the roll-up vortex, a structure strongly related to the boundary layer height, which would otherwise be able to bend around the vane to form a horseshoe vortex. In this case the vane probably acts more like a roughness element, passing the roll-up vortex above itself without significant deformation. Note that although the horseshoe vortex sleeve $H_S$ is not present, the primary vortex $P$ is generated closer to the wall (at the tip of the vane), resulting in a more pronounced secondary structure $S$ which separates and results in the discrete vortex $D$.

For larger $h/\delta$, the horseshoe vortex appears in nearly all regions, independently of vane angle. The secondary structure $S$ generally seems to be more frequently occurring for the lower range of investigated $\beta$. The separated structure $S$ and the pinched off structure $D$ are generally more frequent in the mid- and low-range of vane angles. In the prescribed optimal vane angle range~\cite{Lin2002} about approximately $\beta \approx 15^{\circ} - 25^{\circ}$, the resulting wake is far from the ideal monopole, which, at least in this test bed, appears only for the highest vane angle and height far from the optimum.

\begin{figure}[!ht]
\includegraphics[width=1.0\textwidth]{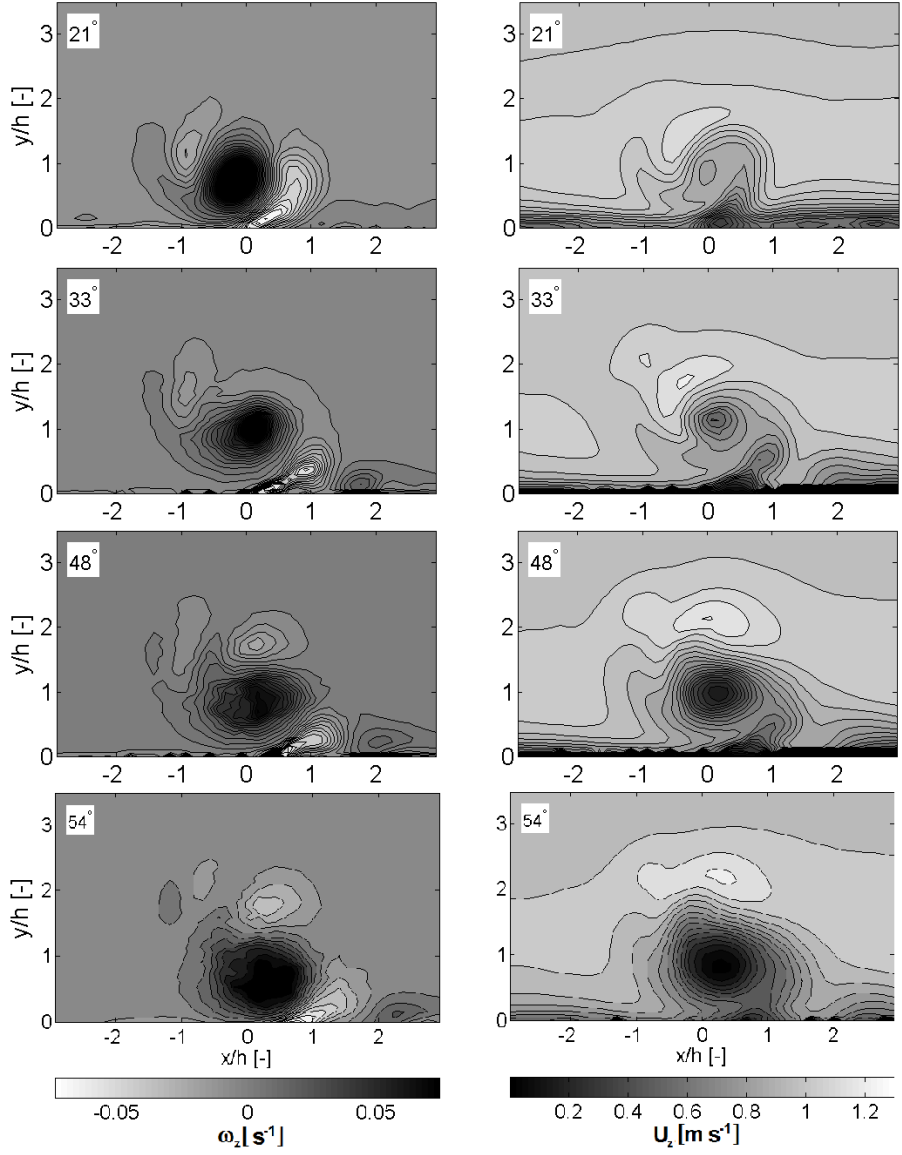}
\caption{Correspondence of structures in the streamwise vorticity (left) and streamwise velocity (right) fields for vane angles $21^{\circ}$, $33^{\circ}$, $48^{\circ}$ and $54^{\circ}$ at $z/h = 3$.}
\label{fig:8}
\end{figure}

\begin{figure}[!ht]
\includegraphics[width=1.0\textwidth]{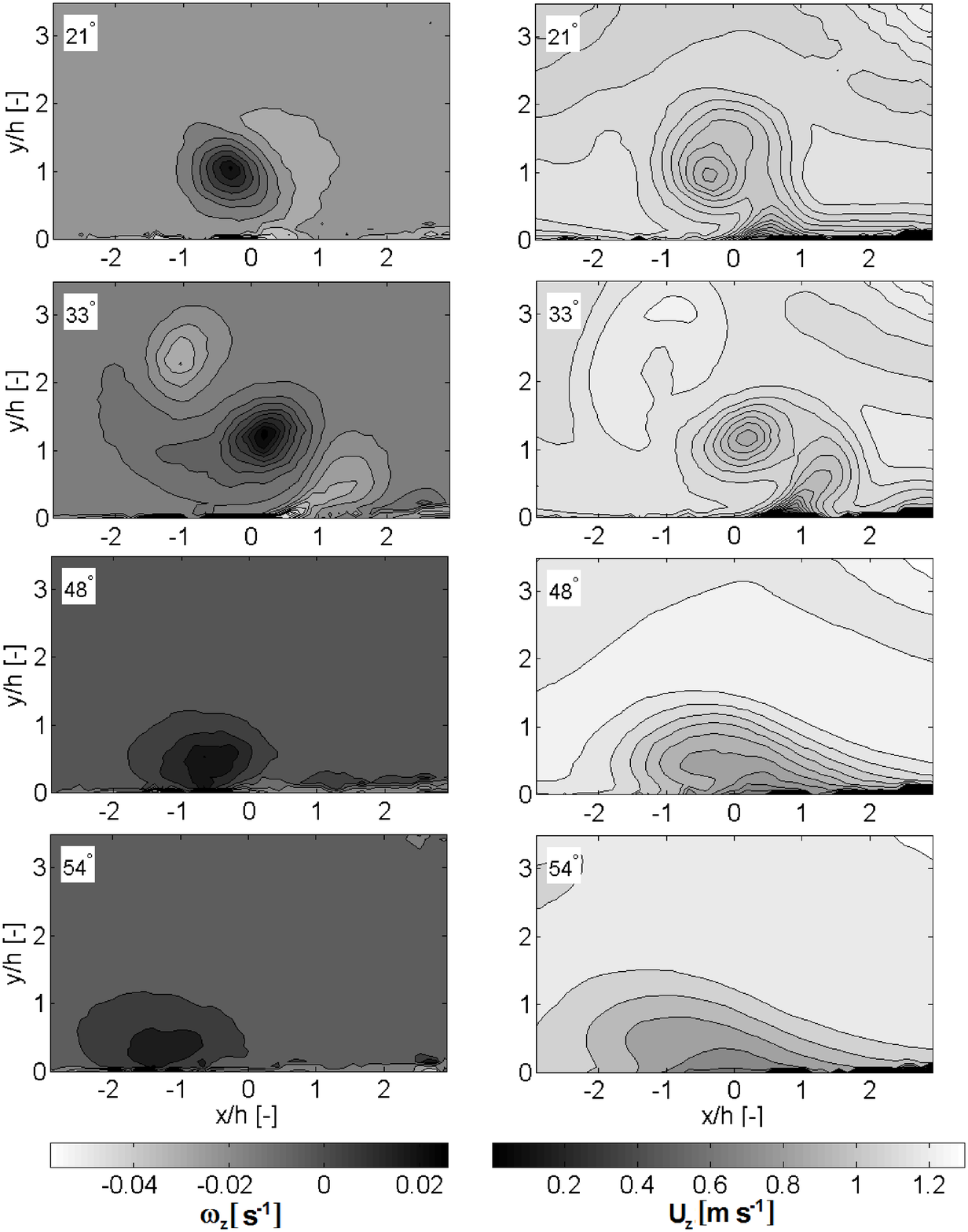}
\caption{Correspondence of structures in the streamwise vorticity (left) and streamwise velocity (right) fields for vane angles $21^{\circ}$, $33^{\circ}$, $48^{\circ}$ and $54^{\circ}$ at $z/h = 10$.}
\label{fig:9}
\end{figure}

More detailed representations of the corresponding averaged measured near and far wake can be seen in Figures~\ref{fig:8} and~\ref{fig:9}. These figures display the streamwise vorticity (left column) and streamwise velocity (right column) components mapped with iso-contours in the near and far wake for the same vane angles as in Figure~\ref{fig:6}. Though unsteady, the average fields of the near wake can be compared in this way due to the linear relation between the velocity and vorticity~\cite{Velteetal2009}. It is clear from the figures that the vorticity and velocity fields have a similar structure, even for structures other than the primary, which is further consistent with the analysis of helical symmetry of the primary vortex of~\cite{Velteetal2009}. This correspondence is a previously unknown property of the complex multiple vortex wakes which is of both fundamental and practical interest, in particular for the construction of more accurate vortex generator wake models.

\section{Vortex state effect on the primary vortex}
\label{sec:Vortexstateeffectontheprimaryvortex}

One may argue that the generation of the lateral adverse pressure gradient by the primary vortex $P$ has a substantial influence on the far wake, since it determines the vorticity flux of opposite rotation generated underneath of $P$. Therefore, it is of interest to study how the far wake states appear in relation to the strength of the primary vortex, $\Gamma_p$, and its center distance to the wall in the far wake, $h_p$, which are assumed to be the two primary factors determining the strength of the adverse lateral pressure gradient.

$h_p$ could be detected directly from the SPIV measurements. Note that even though it is non-trivial to determine the position of the wall in PIV measurements, the camera positions, and therefore the measurement field, remained the same in all measurements and therefore the relative position to the wall was unchanged. An error in the determination of the position of the wall can therefore only contribute with a systematic bias, corresponding to a constant offset of the trends. The circulation of $P$ was found by integration along the vortex core radius, which in turn was determined from a Lamb-Oseen vortex streamwise vorticity distribution in accordance with the methodology employed by~\cite{Velteetal2009}.

\begin{figure}[!ht]
\includegraphics[width=0.8\textwidth]{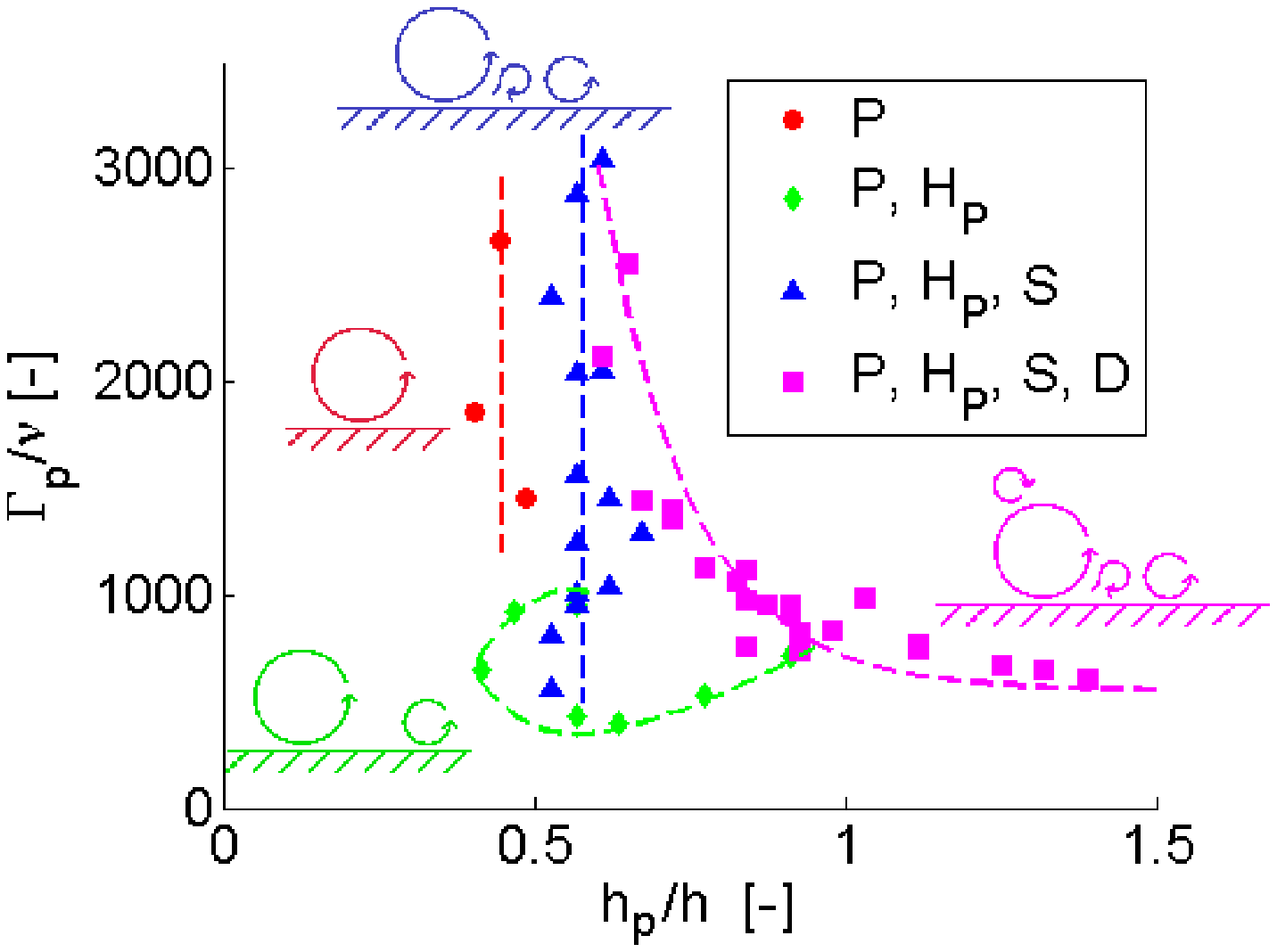}
\caption{Intensity map/bifurcation diagram of the principal far‐wake vortex states as a function of primary vortex strength and primary vortex center distance to the wall.}
\label{fig:10}
\end{figure}

Plotting the far wake measurement results in ($\Gamma_p, h_p$)-space resulted in an intensity map (also called bifurcation diagram for reasons that will be obvious later in the text) displayed in Figure~\ref{fig:10}, where the different topologies are indicated by the following symbols: \textcolor{red}{$\bullet$} only the primary vortex present ($P$), \textcolor{green}{$\blacklozenge$} two vortex regime ($P$ and $H_P$), \textcolor{blue}{$\blacktriangle$} three vortex regime ($P$, $H_P$ and $S$) and \textcolor{magenta}{$\blacksquare$} four vortex regime ($P$, $H_P$, $S$ and $D$). Note that, again, only four of the states presented in Figure~\ref{fig:7} have been plotted since these represent the most frequently occurring states in the parametric study. These states are also depicted in Figure~\ref{fig:10} by figures and indicative lines of the intensity dependence on primary vortex center height are added to the graph, for clarity. The axes are non-dimensionalized by the kinematic viscosity $\nu$ and the vortex generator height, $h$, respectively.

It is clear from the figure that the resulting topology in the stable far wake depends on these main vortex parameters $\Gamma_p$ and $h$. For high values of the primary vortex circulation $\Gamma_p$, the states merge into a narrow band around $h_p/h \approx 0.6$. In this case the primary vortex $P$ is strong enough to dominate the behavior of the vortex system and the wake thus appears stable. For lower values of the primary vortex circulation, the influence of the secondary structures on the primary vortex increases and eventually the band bifurcates into four separate branches, determined by the far wake state.

In the first case where only the primary vortex is present (\textcolor{red}{$\bullet$}), its center-height above the wall is very stable. This system, consisting of a single vortex and its mirror image propelling each other by induction sideways (in the current case to the right) along the wall, is well established. The next branch (\textcolor{blue}{$\blacktriangle$}) with a wake state with two vortices of the same sign ($P$ and $H_P$) seems to follow a complex pattern, which supports the current opinion that a condition for equilibrium of two vortices of the same sign near a wall is a complex problem in vortex dynamics. Note that both of these states, as seen from Figure~\ref{fig:7}, appear outside the range for common applications of vortex generators and seem therefore to be solely of academic interest - at least in the current test bed.

The last two branches are very common in the range of applicable vane heights and angles mentioned earlier according to Figure~\ref{fig:7}. The third branch (\textcolor{blue}{$\blacktriangle$}) has a relatively weak structure $S$ that seems to not affect $P$ considerably, but creates a buffer between $P$ and $H_P$ so that $P$ remains stable as in the first branch. In the last branch (\textcolor{magenta}{$\blacksquare$}), where the separating structure $S$ below the primary vortex becomes stronger, the pinched-off vortex $D$ forms more easily and grows stronger in relation to the primary vortex. This effect is, again, more pronounced for small vanes, where the primary vortex is generated closer to the wall (at the tip of the generator) and for small vane angles. Judging from the observed vorticity maps of the far wakes, $D$ seems to be forming an up-wash pair together with $P$, causing the motion away from the wall.

\begin{figure}[!ht]
\includegraphics[width=0.8\textwidth]{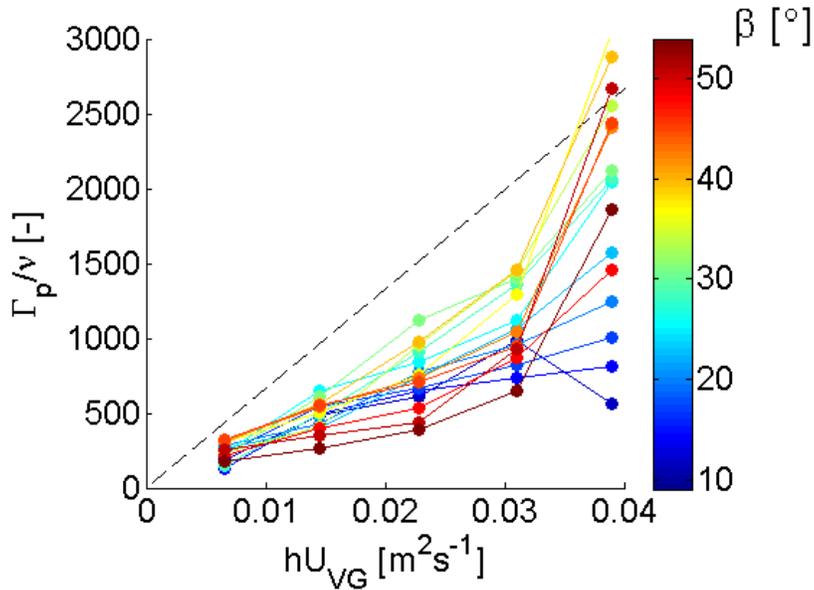}
\caption{Test of linear models based on simple airfoil theory for estimation of circulation of the primary vortex. The dashed line represents an `ideal' vortex generator wake with a linear growth in circulation $\Gamma_p \propto h U_{VG}$, where $U_{VG}$ is the average streamwise velocity of the unperturbed boundary layer at the upper edge of the vortex generator. Each colored line represent the measured primary vortex circulation for a vane angle $\beta$, spanning across the range of investigated vortex generator heights.}
\label{fig:11}
\end{figure}

Despite these observed effects, many workers have been proposing engineering models based on the vortex generator geometry and ideas reminiscent of simple airfoil theory in an attempt to estimate the primary vortex circulation a priori (see, e.g.,~\cite{Pearcey1961,Jones1957,Smith1994,Lamb1932,HansenWestergaard1995,Chaviaropoulos2003}). These models are typically of the form $\Gamma_p \propto h U_{VG}$, where $h$ is the vortex generator height and $U_{VG}$ is the average streamwise velocity at the upper tip of the vortex generator. Note that this type of simplified (inviscid) model assumes the existence of merely a primary vortex.

As a test of these models, Figure~\ref{fig:11} displays the normalized primary vortex circulation, $\Gamma_p/\nu$, as a function of $h U_{VG}$ and for a range of acquired angles $\beta$ (color coded) in the data separately. This figure clearly demonstrates that $\Gamma_p/\nu$ does indeed not necessarily vary linearly with neither the device height $h U_{VG}$, nor with the device angle $\beta$, as expected from simple airfoil theory. On the contrary, the measurements often deviate from the linear trend assumed by simple airfoil theory (e.g., the dashed line Figure~\ref{fig:11}), in particular for the highest values of $h U_{VG}$.

\section{Summary and conclusions}
\label{sec:Summaryandconclusions}

The current work focuses on the recently observed far-wake secondary structures accompanying the classical primary wing-tip vortex. The important application of heat transfer enhancement is of special focus for these results, since the structure of the wake has been shown to potentially have a substantial impact on the resulting heat transfer properties~\cite{Min2010}.

In the work presented, series of Stereoscopic PIV measurements have been conducted in cross-planes in the wake of a single plate-shaped rectangular vortex generator. Due to viscous effects in the near wall region, adding to the complexity of the flow, the resulting vortex system can deviate substantially from the classical expected ideal flow monopole generated at the vortex generator tip~\cite{Taylor1947,Taylor1948a,Taylor1948b,Taylor1950}. From more recent theory and previous experience, the wake was instead expected to contain a basic vortex system (consisting of a primary `wing-tip' vortex and a horseshoe vortex) and a secondary vortex structure washed up by the primary one at the wall, occasionally pinching off a discrete vortex that orbits around the primary one.

Parametric measurements show that the far wake may stabilize in a range of states which often constitute subsets of this combined system. The various regimes detected have been mapped by registration of the wake vorticity in the far wake and it is observed that the near wake is unstable. A regime map displaying the regimes as a function of vane angle and height has been produced. For practical applications only a subset of this regime map in Figure~\ref{fig:7} should be considered, say, $\beta \approx 9^{\circ} - 36^{\circ}$. A range of parametric studies, as presented in the review article by~\cite{Lin2002}, show that commonly the most optimum parameters for the vortex generator is $\beta \approx 14^{\circ} - 30^{\circ}$ and $h/\delta \approx 0.1 - 0.6$ for micro- and regular vortex generators. These reported studies were conducted in widely different flow cases and the optimum parameters therefore seem to be highly dependent on the oncoming flow and the test bed. Note that in the present study, the commonly expected regime with only the primary vortex appears only for the very extreme parameter values, rarely used in applications. Note also that the secondary structures can, if desired, sometimes be avoided in applications by arranging the vortex generators in a cascade to produce counter-rotating vortices. In cases of maximizing wall shear stress, which can be of importance in e.g. separation control applications, the secondary structures appear to cancel out (see, e.g.,~\cite{GodardStanislas2006,Velte2011}).

Though never previously reported for vortex generator induced flows, the basic vortex system is in general well-known and expected in this type of flow situations. It is, however, interesting to study the generation of the secondary vortex propelled by the interaction of the primary vortex and the wall. From earlier work (see, e.g.,~\cite{HarveyPerry1971}) it is seen that this type of structure is generated as a consequence of local separation due to the adverse pressure gradient generated by the primary vortex and the wall. The primary vortex circulation and its distance to the wall influences this adverse pressure gradient and the effects of this are studied. The resulting intensity map (Figure~\ref{fig:10}) reveals that for high values of circulation, all states converge into a narrow band of vortex center heights. As the circulation diminishes, the primary vortex loses its predominance, the secondary structures begin to significantly affect the primary vortex trajectory and the states bifurcate into separate branches. It is evident that the relative strength of the secondary vortex structure to the primary one has a large impact. These results support the hypothesis that the trajectory of the primary vortex may differ considerably from the one predicted by inviscid theory, see, e.g.,~\cite{Pearcey1961,Jones1957,Smith1994,Lamb1932,HansenWestergaard1995,Chaviaropoulos2003}.

Even though the specific results may be unique for the test bed under consideration, it is clear that secondary effects are not necessarily negligible, as has commonly been assumed. In particular, one should be aware, see also~\cite{Min2010}, that these effects can impact significantly the mixing and heat transfer properties of the flow in ways that inviscid models cannot correctly predict.

From the measurements, an additional important fact about the streamwise vorticity field has been established. A correspondence in the structures observed in the streamwise vorticity and the streamwise velocity fields was found. This is especially interesting since the streamwise vorticity is a result of the secondary velocity components normal to the streamwise (flow) direction. This has previously only been established for the primary vortex~\cite{Velteetal2009}, but is in the current study seen to apply for the entire multiple vortex wake. This result is of both fundamental and practical interest, not least since it provides further clues for the construction of more reliable vortex generator wake models.

The correspondence between structures observed in the streamwise velocity and streamwise vorticity is an important result, since it is directly related to vortex generator performance and mixing. From a fundamental point of view, this effect is naturally of interest for connecting the primary and secondary velocity components. From a practical point of view, this knowledge can aid in the construction of future, more accurate, vortex generator wake models.

\section{Acknowledgements}

This work has been carried out with the support of EUDP-2009-II-grant journal no. 64009-0279 and the Danish Council for Strategic Research under the project COMWIND - Center for Computational Wind Turbine Aerodynamics and Atmospheric Turbulence: grant 2104-09-067216/DSF (\url{http://www.comwind.org}) and the Russian Science Foundation (grant no. 14-29-00093).





\end{document}